\documentclass{ws-procs9x6}

\begin{document}

\title{QUANTUM, SPIN AND QED EFFECTS IN PLASMAS}
\author{G. BRODIN$^*$ and M. MARKLUND}

\address{Department of Physics, Ume\aa\ University, SE-901 87 Ume\aa, Sweden \\
$^*$E--mail: gert.brodin@physics.umu.se}

\begin{abstract}
Plasmas are usually described using classical equations. While this is often
a good approximation, where are situations when a quantum description is
motivated. In this paper we will include several quantum effects, ranging
from particle dispersion, which give raise to the so called Bohm potential,
to spin effects, and to quantum electrodynamical effects. The later effects
appears when the field strength approaches the Schwinger critical field,
which may occur in for example astrophysical systems. Examples of how to
model such quantum effects will be presented, and the phenomena resulting
from these models will be discussed.
\end{abstract}

\bodymatter

\section{Introduction}

A characteristic feature of standard plasmas is the domination of collective
forces over single particle forces. This scaling is equivalent to saying
that there should be large number of particles in a Debye sphere. For this
regime, as opposed to strongly coupled plasmas where the opposite condition
holds, classical equations of motion is usually thought to be adequate.
While this certainly applies to some of the wellknown quantum effects, it is
not true in general, however. In this article we will focus on plasmas that
are classical in the sense that there are large numbers of particles in a
Debye sphere, which implies the importance of collective processes. But at
the same time we will include a number of different quantum plasma effects,
that have been of much interest recently see e.g.\ Refs.\ \cite
{collection2A,collection2B,collection2C,collection2D,collection2E,collection2F,collection2G,collection2H,collection2I,collection2J,collection2K,collection2L,Haas-HarrisSheet}%
, starting . The interest in quantum plasma effects has several different
origins, for example recent progress in nanoscale technology \cite{Manfredi}%
, various astrophysical applications \cite
{brodin-marklund-eliasson-shukla,Baring,Beskin-book}, high intensity effects
made relevant by the continuous increase of laser powers \cite
{collectionA,Marklund}, as well as a general theoretical motives \cite
{marklund-brodin,brodin-marklund,probing,brodin-marklund-stenflo,holland}.
As indicated above, the combined focus on collective and quantum plasma
effects are to some extent contradictory, as in many cases these effects are
important in different regimes. Nevertheless there are several important
reasons to treat them simultaneously:

\begin{enumerate}
\item  \textit{Unification:} Before a detailed calculation has been done, it
can be difficult to know whether quantum or collective effects will be
dominant in a specific problem. In this case it is useful to be able to
start from a set of equations including both types of phenomena.

\item  \textit{Different scalings:} Certain quantum effects, in particular
those due quantum electrodynamics (QED) and particle spin does not
necessarily become insignificant even if there is a large number of
particles in a Debye sphere. For such plasmas, collective and quantum
effects can simultaneously be important.

\item  \textit{Symmetry dependent effects: }The symmetry properties of the
standard and quantum terms differ to some extent in the equations of motion.
Thus for a problem with a specific geometry, a classical effect may
sometimes vanish due to a symmetry, while a small quantum effect survives
and dominate the dynamical picture.

\item  \textit{Extreme regimes: }In certain extreme regimes, as for example
found in astrophysics, the formal conditions for collective and quantum
effects to be important simultaneously can be fulfilled.
\end{enumerate}

In this paper we will describe a number of different quantum phenomena that
can be fit into the standard Maxwell-Fluid model by adding various terms to
the classical equations. In particular we will be dealing with particle
dispersion \cite
{collection2A,collection2B,collection2C,collection2D,collection2E,collection2F,collection2G,collection2H,collection2I,collection2J,collection2K,collection2L,Haas-HarrisSheet}%
, giving raise to the so called Bohm potential, effects related to the
spin-half properties of the particles \cite{marklund-brodin,brodin-marklund}%
, as well as QED effects such as vacuum polarization and magnetization \cite
{collectionA,Marklund}, which becomes important for field strengths
approaching the Schwinger critical field, $E_{\text{crit}}\equiv
m_{e}^{2}c^{3}/\hbar e\approx 10^{16}\ \mathrm{V/cm^{-1}.}$ Here $m_{e}$ is
the electron mass, $c$ is the speed of light in vacuum, $e$ is the
elementary charge and $\hbar $ is Plank's constant. The applicability of the
presented models will be discussed, and a number of phenomena induced by the
quantum terms will be shown. Finally, various applications to specific
plasmas will be pointed out.

\section{Particle dispersion and Fermi pressure}

Naturally the most basic quantum effect is that particles are described by
wave functions rather than classical point particles. Following e.g. Ref.\ 
\cite{Manfredi}, the particles are described by the statistical mixture of $%
N $ states $\psi _{i}$, $i=1,2,\dots ,N$ where the index $i$ sums over all
particles independent of species. We then take each $\psi _{i}$ to satisfy a
single particle Schrdinger equation where the potentials $(\mathbf{A},\phi
) $ is due to the collective charge and current densities, i.e. $\varepsilon
_{0}\nabla ^{2}\phi =-\sum_{i=1}^{N}q_{i}p_{i}|\psi _{i}|^{2}$, etc., where $%
p_{i}$ is the occupation probability of state $\psi _{i}$.This model amounts
to assume that all entanglement between particles are neglected. To derive a
fluid description we make the ansatz $\psi _{i}=\sqrt{n_{i}}\exp
(iS_{i}/\hbar )$ where $n_{i}$ is the particle density, $S_{i}$ is real, and
the velocity of the $i$'th particle is $\mathbf{u}_{i}=\nabla
S_{i}/m_{i}-(q_{i}/m_{i}c)\mathbf{A}$. Next we define the global density and
velocity as $n=\sum_{j}p_{j}n_{j}$ and $\mathbf{u}=\sum_{j}p_{j}n_{j}\mathbf{%
u_{j}}/n$, where $j$ runs over all particles. Separating the real and the
imaginary part in the Schrdinger equation, we obtain the continuity
equation 
\begin{equation*}
\frac{\partial n}{\partial t}+\nabla \cdot \left( n\mathbf{u}\right) =0,
\end{equation*}
and the momentum equation 
\begin{equation*}
\frac{\partial \mathbf{u}}{\partial t}+\left( \mathbf{u}\cdot \nabla \right) 
\mathbf{u}=\frac{q}{m}\left( \mathbf{E}+\mathbf{u}\times \mathbf{B}\right) -%
\frac{1}{mn}\nabla P+\frac{\hbar ^{2}}{2m^{2}}\nabla \left( \frac{1}{\sqrt{n}%
}\nabla ^{2}\sqrt{n}\right) .
\end{equation*}
The last term is the gradient of the so called Bohm potential, and the
tendency to smoothen a density profile naturally reflects the dispersive
tendencies of a localized wave packet. Furthermore, we stress that the
pressure term contains both the fermion pressure, $P_{F}$, and the thermal
pressure, $P_{t}$. For low temperature plasmas, where the Fermi pressure is
of most significance, $P_{F}$ can be written as $P_{F}=(4\pi ^{2}\hbar
^{2}/5m)(3/8\pi )^{2/3}n^{5/3}$.

As a simple illustration of some effects due to the quantum terms we can
study linear wave propagation in a homogeneous plasma that may - or may not
- be magnetized. Since both the thermal pressure, the Fermi pressure and the
Bohm potential becomes proportional to $\nabla n_{1}$, where $n_{1}$ is th
density perturbation of the total density $n=n_{0}+n_{1}$, it turns out \cite
{lundin}the above quantum effects can be captured by making the following
simple substitutions 
\begin{equation}
v_{t}^{2}\rightarrow v_{t}^{2}+\frac{3}{5}v_{F}^{2}+\frac{\hbar ^{2}k^{2}}{%
4m^{2}}.  \label{Eq:substition1}
\end{equation}
in any classical linear dispersion relations. Here $v_{t}$ is the thermal
velocity, $v_{F}$ the Fermi velocity and $k$ the wavenumber of the
perturbation. As a specific example we may consider Langmuir waves in which
case the quantum version of the dispersion relation thus becomes 
\begin{equation}
\omega ^{2}=\omega _{p}^{2}+k^{2}\left( v_{t}^{2}+\frac{3}{5}v_{F}^{2}+\frac{%
\hbar ^{2}k^{2}}{4m^{2}}\right)  \label{Eq:DR-Langmuir}
\end{equation}
where $\omega _{p}$ is the plasma frequency. The dispersion relation (\ref
{Eq:DR-Langmuir}) was recently experimentally verified in X-ray scattering
experiments made in Laser produced plasmas \cite{Langmuir}.

\section{Particle spin}

The treatment of the previous section can be generalized to include the
effects of particle spin. The following modifications are then necessary 
\cite{marklund-brodin,brodin-marklund}:

\begin{enumerate}
\item  \bigskip Replace the Schrdinger equation for a scalar wave function
with the Pauli equation for the spinors.

\item  Decompose the spinors according to $\psi _{i}=\sqrt{n_{i}}\exp
(iS_{i}/\hbar )\varphi _{i}$, where $\varphi _{i}$ is a normalized two
spinor.

\item  Introduce the velocity $\mathbf{u}_{i}$ and the spin vector $\mathbf{s%
}_{i}$ as $\mathbf{u}_{i}=(1/m)(\nabla S_{i}-i\hbar \varphi _{i}^{\dag
}\nabla \varphi _{i})-(q_{i}/m_{i}c)\mathbf{A}$ and $\mathbf{s}_{i}=(\hbar
/2)\varphi _{i}^{\dag }\sigma \varphi _{i}$, where $\sigma =(\sigma
_{1},\sigma _{2},\sigma _{3})$ and $\sigma _{1,2,3}$ are the Pauli spin
matrices.
\end{enumerate}

It is no surprise that the resulting equations are considerably more
complicated than the spinless equations in the preceding section. Rather
than presenting the full theory (see Refs. \cite
{marklund-brodin,brodin-marklund}) here, we will focus on the leading
contributions where a number of terms of higher order in $\hbar $ are
neglected. The spin effects can then be captured by a spin force $\mathbf{F}%
_{sp}$ that is added to the momentum equation, and a magnetization current%
\textbf{\ }$\mathbf{j}_{m}$ associated with the spins. These expressions in
turn depend on a macroscopic spin vector $\mathbf{s=}\sum_{i}\mathbf{s}_{i}/n
$, that is described by a separate evolution equation complementing the
Maxwell-fluid system. The results for the electrons, denoted by index $e$
are 
\begin{eqnarray}
\mathbf{F}_{sp} &=&\frac{2\mu _{B}n_{e}}{\hbar }s^{j}\nabla B_{j}
\label{Eq:spin-force} \\
\mathbf{j}_{m} &=&\nabla \times \mathbf{M=}\nabla \times \left( \frac{%
2n_{e}\mu _{B}\mathbf{s}}{\hbar }\right)   \label{Eq:Magnetization-current}
\\
\left( \frac{\partial }{\partial t}+\mathbf{u\cdot }\nabla \right) \mathbf{s}
&=&\frac{2\mu _{B}}{\hbar }\mathbf{B}\times \mathbf{s}
\label{Eq:Spin-evolution}
\end{eqnarray}
where $\mu _{B}=e\hbar /2m$ is the Bohr magneton, $\mathbf{B}$ is the
magnetic field, $\mathbf{M}$ is the magnetization vector and we use the
Einstein summation convention in Eq. (\ref{Eq:spin-force}). The spin effects
associated with the ions is usually smaller due to their larger mass. For a
generalization including the spin contribution for an arbitrary particle
species, see Refs. \cite{marklund-brodin,brodin-marklund}.

There is a rich variety of new dynamical effects associated with the spins,
as described by Eqs. (\ref{Eq:spin-force})-(\ref{Eq:Spin-evolution}).
However, in order to start exploring the dynamics, we must first have an
expression for the spin vector in thermodynamic equilibrium. The result for
spin half particles is \cite{marklund-brodin,brodin-marklund} $\mathbf{s=}%
(\hbar /2)\tanh (\mu _{B}B_{0}/T)$, where $B_{0}$ is the unperturbed
magnetic field and $T$ is the temperature given in energy units. A simple
example of the results that can be derived from the Maxwell-Fluid results
complemented by (\ref{Eq:spin-force})-(\ref{Eq:Spin-evolution}) is the
modification of the Alfvn velocity. Taking the MHD limit, it turns out that
the Alfvn velocity $C_{A}=\left( B_{0}^{2}/\mu _{0}\rho _{0}\right) ^{1/2}$
is modified according to \cite{rapid-spin} 
\begin{equation}
C_{A}\rightarrow \frac{C_{A}}{(1+(\hbar \omega _{pe}^{2}/2mc^{2}\omega
_{ce}^{(0)})\tanh (\mu _{B}B_{0}/T))^{1/2}}  \label{Eq:Alfven-subst}
\end{equation}
where $\omega _{ce}^{(0)}$ $=eB_{0\mathrm{ext}}/m$is the electron cyclotron
frequency due to the external field $B_{0\mathrm{ext}}$ only, i.e. the
contribution from the zero order spin magnetization is excluded. The
substitution (\ref{Eq:Alfven-subst}) applies both for the shear Alfvn mode
as well as for the fast and slow magnetosonic modes.

\section{High field and short wavelength QED effects}

The first order QED effects can effectively be modeled through the
Heisenberg-Euler Lagrangian density \cite{Heisenberg-Euler, Schwinger}. This
Lagrangian describes a vacuum perturbed by a slowly varying electromagnetic
field. The effect of rapidly varying fields can be accounted for by adding a
derivative correction to the Lagrangian \cite{Mamaev-1981}. This correction
is referred to as the derivative QED correction or the short wavelength QED
correction. The Heisenberg-Euler Lagrangian density with the derivative
correction reads 
\begin{eqnarray}
&&\!\!\!\!\!\!\mathcal{L}=\mathcal{L}_{0}+\mathcal{L}_{HE}+\mathcal{L}_{D} 
\notag \\
&&\!\!\!\!=\frac{\varepsilon _{0}}{4}F_{ab}F^{ab}+\frac{\varepsilon
_{0}^{2}\kappa }{16}\left[ 4\left( F_{ab}F^{ab}\right) ^{2}+7\left( F_{ab}%
\widehat{F}^{ab}\right) ^{2}\right] +\sigma \varepsilon _{0}\left[ \left(
\partial _{a}F^{ab}\right) \left( \partial _{c}F_{\phantom{b}b}^{c}\right)
-F_{ab}\Box F^{ab}\right] ,
\end{eqnarray}
where $\mathcal{L}_{0}$ is the classical Lagrangian density, while $\mathcal{%
L}_{HE}$ represents the Heisenberg-Euler correction due to first order
strong field QED effects, $\mathcal{L}_{D}$ is the derivative correction, $%
\Box =\partial _{a}\partial ^{a}$ is the d'Alembertian, $F^{ab}$ is the
electromagnetic field tensor and $\widehat{F}^{ab}=\epsilon ^{abcd}F_{cd}/2$
where $\epsilon ^{abcd}$ is the totally antisymmetric tensor. The parameter $%
\kappa =2\alpha ^{2}\hbar ^{3}/45m^{4}c^{5}$ gives the nonlinear coupling, $%
\sigma =(2/15)\alpha c^{2}/\omega _{e}^{2}$ is the coefficient of the
derivative correction and $\alpha =e^{2}/4\pi \hbar c\varepsilon _{0}$ is
the fine structure constant, where $\varepsilon _{0}$ is the free space
permittivity. We obtain the field equations from the Euler-Lagrange
equations $\partial _{b}\left[ \partial \mathcal{L}/\partial F_{ab}\right]
=\mu _{0}j^{a}$, 
\begin{equation}
\left( 1+2\sigma \Box \right) \partial _{a}F^{ab}=2\varepsilon _{0}\kappa
\partial _{a}\left[ \left( F_{cd}F^{cd}\right) F^{ab}+\tfrac{7}{4}\left(
F_{cd}\widehat{F}^{cd}\right) \widehat{F}^{ab}\right] +\mu _{0}j^{b},
\end{equation}
where $j^{a}$ is the four-current and $\mu _{0}$ is the free space
permeability.

The corresponding sourced Maxwell equations resulting from the derivative
corrected field equation then become 
\begin{equation}
\left[ 1+2\sigma \left( -\frac{1}{c^{2}}\frac{\partial ^{2}}{\partial t^{2}}%
+\nabla ^{2}\right) \right] \nabla \cdot \mathbf{E}=\frac{\rho +\rho _{%
\mathrm{vac}}}{\varepsilon _{0}},
\end{equation}
\begin{equation}
\left[ 1+2\sigma \left( -\frac{1}{c^{2}}\frac{\partial ^{2}}{\partial t^{2}}%
+\nabla ^{2}\right) \right] \left( \nabla \times \mathbf{B}-\frac{1}{c^{2}}%
\frac{\partial \mathbf{E}}{\partial t}\right) =\mu _{0}(\mathbf{j+j}_{%
\mathrm{vac}}),
\end{equation}
where the vacuum charge density is $\rho _{\mathrm{vac}}=-\nabla \cdot 
\mathbf{P}$ and the vacuum current density is $\mathbf{j}_{\mathrm{vac}%
}=\partial \mathbf{P}/\partial t-\nabla \times \mathbf{M}$ with the vacuum
polarization and magnetization given by 
\begin{eqnarray}
\mathbf{P} &=&2\epsilon _{0}^{2}\kappa \lbrack 2(E^{2}-c^{2}B^{2})\mathbf{E}%
+7c^{2}(\mathbf{E}\cdot \mathbf{B})\mathbf{B}]  \label{Eq:Polarization} \\
\mathbf{M} &=&2c^{2}\epsilon _{0}^{2}\kappa \lbrack -2(E^{2}-c^{2}B^{2})%
\mathbf{B}+7(\mathbf{E}\cdot \mathbf{B})\mathbf{E}]  \label{Eq:Magnetization}
\end{eqnarray}
respectively. The source free Maxwell equations are $\nabla \cdot \mathbf{B}%
=0$ and 
\begin{equation}
\nabla \times \mathbf{E}=-\frac{\partial \mathbf{B}}{\partial t}.
\end{equation}
The QED vacuum contribution can give raise to a large number of physical
effects \cite{Marklund}. For example the nonlinear vacuum terms implies
processes such as photon-photon scattering. However, in order to keep our
examples simple, we here consider just linear wave propagation, and also
treat the high field effects proportional to $\kappa $ and the short
wavelength effects proportional to $\sigma $ separately from now on.

As our first example we consider short wavelength linear wave propagation in
a magnetized plasma, and keep only the QED terms proportional to $\sigma $.
It turns out that the effects due to a finite $\sigma $ can be included in a
very simple manner, by making the substitution 
\begin{equation}
\omega _{ps}^{2}\rightarrow \frac{\omega _{ps}^{2}}{1-\zeta }.
\end{equation}
everywhere in the susceptibility tensor of a plasma \cite{lundin}, where $%
\zeta =2\sigma (\omega ^{2}/c^{2}-k^{2})$. In most cases the short
wavelength QED corrections is a very small effect. The possibility to
confirm such effects in laboratory has been discussed in some detail by Ref. 
\cite{lundin}.

As our second QED example we consider the effects of the vacuum polarization
a magnetization due to a strong external magnetic field $\mathbf{B}=B_{0}%
\widehat{\mathbf{z}}$. We note that the term proportional to $\kappa $
contributes with terms that are linear in the wave field and quadratic in $%
B_{0}$. Following Ref.\ \cite{Brodin} we find that the susceptibility tensor
can be modified to include the QED effect of strong magnetic fields by
adding the correction 
\begin{equation}
\chi _{\text{{\tiny {QED}}}}=-4\xi \left( 
\begin{array}{ccc}
1-n_{\Vert }^{2} & 0 & n_{\bot }n_{\Vert } \\ 
0 & 1-n^{2}-2n_{\bot } & 0 \\ 
n_{\bot }n_{\Vert } & 0 & -\tfrac{5}{2}-n_{\bot }^{2}
\end{array}
\right) ,
\end{equation}
where $\xi \equiv \kappa \varepsilon _{0}c^{2}B_{0}^{2}=(\alpha /90\pi
)(cB_{0}/E_{\text{crit}})^{2}$, $n_{\bot }=k_{\bot }c/\omega $, $n_{\Vert
}=k_{\Vert }c/\omega $ and $n=kc/\omega $. Here the indices $\bot $ and $%
\Vert $ denote the directions perpendicular and parallel to the external
magnetic field respectively, and we have chosen the wavevector to lie in the 
$xz$-plane. In magnetar environments with extreme magnetic fields, the
parameter $\xi $ can approach unity \cite{Beskin-book}. As a consequence,
wave propagation in the electron-positron plasma surrounding magnetars are
likely to be significantly affected by strong field QED effects.

\section{Concluding remarks}

In this paper we have given a brief review of how the plasma dynamics is
modified by various quantum, spin and QED effects. The approach has been to
modify the Maxwell-Fluid equations, in order to keep contact with the
theoretical results and methods developed for classical systems. In order to
illustrate the usefulness of the modified equations, we have presented some
simple results for linear wave propagation. It should be stressed, however,
that much of the work dealing with quantum and QED effects focuses on
nonlinear phenomena \cite{Marklund}. The set of plasmas where the new
phenomena tend to be important can be briefly described as follows:

\begin{enumerate}
\item  \textit{The Bohm potential and Fermi pressure: }Low temperature
and/or high density plasmas. This includes solid state plasmas, white dwarf
stars and to some extent laser produced plasmas. Ultra-cold plasmas
generated from Rydberg states are also of interest in this context.

\item  \textit{Spin effects: }Due to the complexity of the spin dynamics, it
is difficult to give simple conditions when these effects are important.
However, a few simple rules of thumb can be given: Spin effects are
important if the energy difference between the two spin states is larger
than the thermal energy. This applies to plasmas in the vicinity of
magnetars, and possibly also to ultracold plasmas. Furthermore, spin effects
can be important in low-temperature high density plasmas, similarly to the
ones described in point 1. Finally spin effects can be important if $%
C_{A}^{2}\lesssim \mu _{B}B_{0}/m_{i}$, which contrary to the first
conditions tend to be fulfilled in plasmas embedded in a rather weak
external magnetic field.

\item  \textit{High field QED effects:} The characteristic scale for this
phenomena is the Schwinger critical field. However, since qualitatively new
phenomena (i.e. photon-photon scattering in vacuum) occur due to these
terms, there is a hope to see such QED effects even before the laser
intensities reach this extreme scale. Furthermore, astrophysical plasmas in
the vicinity of magnetars can be subject to magnetic field strengths
exceeding the critical field.

\item  \textit{Short wavelength QED effects:} These are important when the
wavelength approach the Compton wavelength, provided the plasma density is
very high, such that the $c/\omega _{p}$ is comparable or smaller than the
Compton wavelength.
\end{enumerate}

The above list should not be taken too literally, as quantum effects
certainly can be more important whenever a classical effect vanishes due to
some symmetry obeyed by the classical terms only. We conclude this paper by
pointing out that the field of quantum plasmas is a very rich one, and that
many important aspects remain to be discovered.

\end{document}